\documentclass[a4paper,doc,floatsintext,natbib]{apa6}\usepackage[]{graphicx}\usepackage[]{xcolor}
\makeatletter
\def\maxwidth{ %
  \ifdim\Gin@nat@width>\linewidth
    \linewidth
  \else
    \Gin@nat@width
  \fi
}
\makeatother

\definecolor{fgcolor}{rgb}{0.345, 0.345, 0.345}

\usepackage{framed}
\makeatletter
 {\par\unskip\endMakeFramed%
 \at@end@of@kframe}
\makeatother

\definecolor{shadecolor}{rgb}{.97, .97, .97}
\definecolor{messagecolor}{rgb}{0, 0, 0}
\definecolor{warningcolor}{rgb}{1, 0, 1}
\definecolor{errorcolor}{rgb}{1, 0, 0}
\newenvironment{knitrout}{}{} 

\usepackage{alltt}

\usepackage[english]{babel}
\usepackage[utf8x]{inputenc}
\usepackage{amsmath}
\usepackage{graphicx}
\usepackage[colorinlistoftodos]{todonotes}
\usepackage{hyperref}
\usepackage{alltt}  
\usepackage{framed} 
\usepackage{lineno}
\usepackage{color}
\usepackage{natbib}
\usepackage{algorithm}     
\usepackage{algpseudocode} 

\newcommand{\pathBIB}{./bib}

\title{Supervised dimensionality reduction for multiple imputation by chained equations}
\shorttitle{[PREPRINT] SDR for MICE}
\threeauthors
{Edoardo Costantini}
{Kyle M. Lang}
{Klaas Sijtsma}
\threeaffiliations
{Tilburg University, Department of Methodology and Statistics}
{Utrecht University, Department of Methodology and Statistics}
{Tilburg University, Department of Methodology and Statistics}
\authornote{
  Corresponding author's email address: \\
  e.costantini@tilburguniversity.edu; \\
}

\abstract{
Multivariate imputation by chained equations (MICE) is one of the most popular approaches to address missing values in a data set. 
This approach requires specifying a univariate imputation model for every variable under imputation.
The specification of which predictors should be included in these univariate imputation models can be a daunting task. 
Principal component analysis (PCA) can simplify this process by replacing all of the potential imputation model predictors with a few components summarizing their variance.
In this article, we extend the use of PCA with MICE to include a supervised aspect whereby information from the variables under imputation is incorporated into the principal component estimation.
We conducted an extensive simulation study to assess the statistical properties of MICE with different versions of supervised dimensionality reduction and we compared them with the use of classical unsupervised PCA as a simpler dimensionality reduction technique.}

\IfFileExists{upquote.sty}{\usepackage{upquote}}{}
\begin{document}
\maketitle

\setcounter{secnumdepth}{3} 


\section{Introduction}\label{sec:introduction}

Multiple Imputation (MI) is a state-of-the-art missing data treatment in today's data analysis (\citeauthor{schaferGraham:2002}, \citeyear{schaferGraham:2002}; \citeauthor{vanBuuren:2018}, \citeyear{vanBuuren:2018}, p. 30).
Multiple imputations is often implemented through the multivariate imputation by chained equations approach \citep[MICE,][]{vanBuurenOudshoorn:2000}, which can accommodate a wide range of data measurement levels.
This flexibility comes from the possibility of modeling the multivariate joint density of the variables with missing values through a collection of conditional densities for every variable under imputation.

MICE requires specifying a different univariate imputation model for each variable under imputation, which entails deciding on the imputation model form and which predictors to use.
The first decision is usually guided by the measurement level of the variables under imputation.
For example, continuous variables can be imputed using a linear regression model, while binary variables can be imputed using logistic regression.
The second decision concerns which and how many predictors to include in the imputation models and therefore it is more difficult.
The general recommendation has been to follow an inclusive strategy \citep[][]{collinsEtAl:2001}, meaning that as many predictors as possible should be included in the imputation models.
Using as much information as possible from the data leads to multiple imputations that have minimal bias and maximal efficiency~\citep{meng:1994, collinsEtAl:2001}.
Furthermore, including more predictors in the imputation models makes the missing at random assumption (MAR) more plausible~\citep[p. 339]{collinsEtAl:2001}.
Finally, if the imputation model omits variables that are part of the analysis model fitted to the data after imputation, the parameter estimates might be biased \citep[p. 229]{enders:2010} and estimated confidence intervals might be too wide \citep[p. 218]{littleRubin:2002}.
As a result, including more predictors in the imputation models increases the range of analysis models that can be estimated with a given set of imputations~\citep{meng:1994}.

Despite its advantages, the inclusive strategy easily results in singularity issues~\citep[][p. 46]{hastieEtAl:2009} when estimating imputation models.
Consequently, researchers performing MI often face difficult choices on how many and which variables to use as predictors.
High-dimensional prediction models offer an opportunity to specify the imputation models automatically and to include more predictors than traditionally possible.
For example, ridge regression~\citep{hoerlKennard:1970} can estimate regression models with hundreds of predictors; lasso regression~\citep{tibshirani:1996} can perform data-driven variable selection; 
decision trees~\citep[e.g.,][]{breiman:2001} can consider hundreds of variables for their splitting rules; and principal component analysis~\citep[PCA,][]{jolliffe:2002} can summarize a large set of predictors using a few independent linear combinations.

All of these modeling strategies have been implemented in combination with MICE~\citep{zhaoLong:2016, dengEtAl:2016, burgetteReiter:2010, dooveEtAl:2014, shahEtAl:2014, howardEtAl:2015}.
\cite{costantiniEtAl:2023b} compared their performance in terms of estimation bias and confidence interval coverage when applied to data with missing values.
They found that using PCA to create summaries of the many possible imputation model predictors performs particularly well.
In a follow-up study,~\cite{costantiniEtAl:2023} explored different ways of using PCA with the MICE algorithm and found that updating the principal components (PCs) for the imputation of every variable at every iteration provided the lowest bias and the lowest deviation from nominal confidence interval coverage.
However, these results relied heavily on the number of components computed.
With their simulation study,~\cite{costantiniEtAl:2023} showed that to achieve small bias and satisfactory coverage, a researcher imputing the data using PCA to aid imputation model specification should retain at least as many PCs as the number of latent variables in the data generating model.
This is undesirable as researchers usually do not know the true number of latent variables.

PCA is an unsupervised dimensionality reduction technique that summarizes the variability of a set of $P$ variables \{$\mathbf{x}_1, \dots, \mathbf{x}_p$\} measured on $N$ observations with a set of $Q$ PCs, with $Q < P$.
The PCs can be used in any regression model as a replacement for the original predictors, an approach known as Principal Component Regression~\cite[PCR; ][pp. 168-173]{jolliffe:2002}.
PCR addresses possible multicollinearity issues afflicting the model.
However, the PCs obtained by PCR cannot take into account variables that are not part of the set \{$\mathbf{x}_1, \dots, \mathbf{x}_p$\}, a feature that can result in PCs that are unrelated or only weakly related to the dependent variable, which by definition is not included in the set of predictors.
Contrary to PCA, supervised dimensionality reduction (SDR) techniques use the outcome variable to guide the computation so that the resulting PCs are both good representations of the predictor variables and strongly associated with the dependent variable~\citep[e.g.,][]{wold:1975, deJongKiers:1992, bairEtAl:2006}.
Using SDR within MICE might relax the need to know the number of latent variables in the data-generating model described by~\cite{costantiniEtAl:2023} for PCA.
The purpose of this study is to evaluate how SDR techniques can improve upon unsupervised PCR as a univariate imputation model in MICE.

In this study, we considered two questions.
First, what are the statistical properties (bias, coverage, confidence interval width) of parameters estimated from data treated with the MICE algorithm using different versions of SDR as the univariate imputation models?
Second, can using SDR in MICE relax the PCA requirement of using at least as many PCs as the number of latent variables in the data-generating model?
We used a Monte Carlo simulation study to explore the performance of different versions of SDR with the MICE algorithm.

The article is structured as follows.
In Section~\ref{sec:methods}, we describe the MICE algorithm, unsupervised and supervised dimensionality reduction, different versions of SDR, and we propose uses of SDR as a univariate imputation method for MICE.
In Section~\ref{sec:sim-study}, we describe the Monte Carlo simulation study.
Next, we discuss the main findings (Section~\ref{sec:discussion}), our ideas for future research directions (Section~\ref{sec:limits}), and we provide concluding remarks (Section~\ref{sec:conclusions}).

\section{Imputation methods and algorithms}\label{sec:methods}

We use the following notation.
Indices and scalars are denoted by lowercase and uppercase letters.
For example, $i$ is an index enumerating iterations out of $I$ total iterations $(i \in \{1, \dots, I\})$.
Vectors are written in bold lowercase while matrices are denoted by bold uppercase letters.
The superscript $'$ defines the transpose of a matrix.
We use the subscript $obs$ and $mis$ to refer to the observed and missing elements in a vector or matrix.

\subsection{Multivariate imputation by chained equations}\label{subsec:multiple-imputation-chained-equations}

Consider data set $\mathbf{Z}$ with $N$ rows and $P$ columns $\mathbf{z}_{1}, \dots, \mathbf{z}_{P}$.
We assume that $\mathbf{Z}$ is a random draw from a multivariate distribution $f(\mathbf{Z}|\mathbf{\theta})$, where $\mathbf{\theta}$ is a vector of unknown parameters that completely specifies its multivariate distribution.
Let the first $J$ columns of $\mathbf{Z}$ have missing values.
MICE is an iterative algorithm for imputing multivariate missing data on a variable-by-variable basis.
It obtains multiple imputations for the missing values by drawing from the variable-specific conditional distributions of the form:
\begin{equation} \label{eq:uim}
    f(\mathbf{z}_j|\mathbf{Z}_{-j}, \mathbf{\theta}_j),
\end{equation}
where  $\mathbf{z}_j$ is a partially observed variable, $\mathbf{Z}_{-j}$ is the collection of variables in $\mathbf{Z}$ excluding $\mathbf{z}_{j}$, and $\mathbf{\theta}_j$ is a vector of model parameters.
The parameters $\mathbf{\theta}_j$ are specific to the respective conditional distributions and might not determine the unique \emph{true} joint distribution $f(\mathbf{Z}|\mathbf{\theta})$.
We refer to the conditional distributions in Equation~\ref{eq:uim} as (univariate) imputation models.

The MICE algorithm starts by replacing the missing values in each $\mathbf{z}_j$ with initial guesses (e.g., random draws from the observed values).
At iteration $i$, the MICE algorithm imputes successively variables $\mathbf{z}_1$ to $\mathbf{z}_J$ by taking draws from the following distributions:
\begin{align}
    \mathbf{\theta}_j^{(i)} & \sim f(\mathbf{\theta}_j|\mathbf{Z}_{j, \text{obs}}, \mathbf{Z}_{-j}^{(i)}) \label{eq:post_dist},     \\
    \mathbf{z}^{(i)}_{j, \text{mis}}   & \sim f(\mathbf{z}_{j, \text{mis}}|\mathbf{Z}_{-j}^{(i)}, \mathbf{\theta}_j^{(i)})                \label{eq:pred_dist}
\end{align}
Equation~\ref{eq:post_dist} is the fully conditional posterior distribution defined by the product of an uninformative prior distribution for $\mathbf{\theta}_j$ and the likelihood of observing $\mathbf{z}_{j, \text{obs}}$ under the imputation model for $\mathbf{z}_j$.
Equation~\ref{eq:pred_dist} is the posterior predictive distribution from which updates of the imputations are drawn.
In both equations, $\mathbf{Z}_{-j}^{(i)}$ is $(\mathbf{z}_1^{(i)}, \dots, \mathbf{z}_{j-1}^{(i)}, \mathbf{z}_{j+1}^{(i-1)}, \dots, \mathbf{z}_{J}^{(i-1)}, \mathbf{z}_{J+1}, \dots, \mathbf{z}_{P})$, meaning that at all times the most recently imputed values of all variables are used to impute other variables.

After repeating the sampling steps described by Equation \ref{eq:post_dist} and  \ref{eq:pred_dist} for every variable under imputation, the algorithm moves to the next iteration and repeats the same sampling steps for all variables under imputation.
The convergence of the algorithm is usually assessed by plotting the trends of the average imputations across iterations for different starting values.
After convergence, the imputations are assumed to be samples from the target multivariate distribution.
With this process, one can generate as many imputed data sets as desired.
Finally, the analysis model used to answer a substantive researcher question is estimated on each imputed data set, and the parameter estimates are pooled using Rubin's rules \citep{rubin:1987}.

For small values of $P$, the researcher imputing the data can use all of the columns in $\mathbf{Z}_{-j}$ as predictors in the univariate imputation model for $\mathbf{z}_{j}$.
As $P$ grows larger, the imputer needs to decide which predictors to include and which to leave out, a task that can require a considerable amount of expertise in both statistical modeling techniques and the field of the substantive research question.
By summarizing the information in all of the possible predictors with a few linear combinations of the columns of $\mathbf{Z}_{-j}$, PCA and other dimensionality reduction techniques provide an accessible, data-driven way of specifying the imputation models.

\subsection{Principal component analysis}\label{subsec:principal-component-analysis}

PCA is a dimensionality reduction technique that finds a low-dimensional representation of the variables contained in an $N \times P$ data matrix $\mathbf{X}$ with minimal loss of information.
It does so by finding a $P \times Q$ matrix of weights $\mathbf{W}$ that defines $Q$ independent linear combinations of the columns of $\mathbf{X}$\footnote{We follow the common practice of assuming that the columns of $\mathbf{X}$ are mean-centered and scaled to have a variance of 1.} with maximum variance (with $Q \leq P$).
These weights project the columns of $\mathbf{X}$ onto a lower-dimensional subspace to produce the $N \times Q$ matrix $\mathbf{T} = \mathbf{X}\mathbf{W}$ that summarizes the information in $\mathbf{X}$.

The $Q$ columns of $\mathbf{T}$ are called the principal components (PCs) of $\mathbf{X}$.
The first PC of $\mathbf{X}$ is the linear combination of the columns of $\mathbf{X}$ with the largest variance:
\begin{equation} \label{eq:z1}
    \mathbf{t}_1 = \mathbf{x}_1 w_{11} + \mathbf{x}_2 w_{12} + \dots + \mathbf{x}_{P} w_{1P} = \mathbf{X} \mathbf{w}_{1},
\end{equation}
with $\mathbf{w}_1$ being the $P \times 1$ vector of weights that comprises the first column of $\mathbf{W}$.
The second principal component ($\mathbf{t}_2$) is defined by finding the vector of weights $\mathbf{w}_2$ giving the linear combination of $\mathbf{x}_1, \dots, \mathbf{x}_{P}$ with maximal variance out of all the linear combinations that are uncorrelated with $\mathbf{t}_1$.
Every subsequent column of $\mathbf{T}$ can be understood in the same way: for example, $\mathbf{t}_3$ is the linear combinations of $\mathbf{x}_1, \dots, \mathbf{x}_{P}$ that has maximal variance out of all the linear combinations that are uncorrelated with $\mathbf{t}_1$ and $\mathbf{t}_2$.
As a result, all PCs are uncorrelated by definition and every subsequent PC has a lower variance than the preceding one.

We can also think of PCA as the process of projecting the original data from a set of oblique vectors in a $P$-dimensional space to a set of orthogonal vectors in a Q-dimensional subspace.
The weight vectors $\mathbf{w}_1, \dots, \mathbf{w}_{Q}$ define the directions in which the $N$ observations in $\mathbf{x}_1, \dots, \mathbf{x}_{P}$ are projected.
The projected values are the principal component scores $\mathbf{T}$.

Estimating PCA is the task of minimizing the criterion:
\begin{align}
    \begin{split} \label{eq:pca}
        (\mathbf{W}, \mathbf{P}) & = \underset{\mathbf{W}, \mathbf{P}}{\operatorname{argmin}} \; \lVert \mathbf{X} - \mathbf{XWP}' \rVert^2 \\
        & = \underset{\mathbf{W}, \mathbf{P}}{\operatorname{argmin}} \; \lVert \mathbf{X} - \mathbf{TP}' \rVert^2
      \end{split}
\end{align}
subject to the constraint $\mathbf{P}' \mathbf{P} = \mathbf{I}$, where $\mathbf{P}'$ provides the weights for estimating $\mathbf{X}$ from $\mathbf{T}$.
In other words, PCA finds the matrices $\mathbf{W}$ and $\mathbf{P}$ that minimize the reconstruction error $\mathbf{E}_{\mathbf{X}} = \mathbf{X} - \mathbf{XWP}'$.

\subsection{Principal component regression}\label{subsec:principal-component-regression}

PCR replaces the $P$ predictors of a regression model with $Q$ PCs extracted from those predictors.
Given an outcome variable $\mathbf{y}$ and a set of $P$ predictors $\mathbf{X}$, consider a standard regression model:
\begin{equation} \label{eq:lm}
    \mathbf{y} = \mathbf{X}\mathbf{\beta} + \epsilon,
\end{equation}
where $\mathbf{\beta}$ is a $P \times 1$ vector of regression coefficients, and $\epsilon$ is a $N \times 1$ vector of independent normally distributed errors.
With PCR we use $Q$ PCs of $\mathbf{X}$ in its place so that Equation~\ref{eq:lm} can be rewritten as:
\begin{equation} \label{eq:PCR}
    \mathbf{y} = \mathbf{T}\mathbf{\gamma} + \epsilon,
\end{equation}
where $\mathbf{\gamma}$ is a $Q \times 1$ vector of regression coefficients.
The lower dimensionality of $\mathbf{T}$ compared to $\mathbf{X}$ and the independence of its columns allow Equation~\ref{eq:PCR} to address the computational limitations of Equation~\ref{eq:lm} when $P$ is large or when the variables in $\mathbf{X}$ are highly correlated.

The optimal number of components $Q$ is never certain.
A crucial feature of PCA is that the first PC explains the maximum amount of variance in all $P$ predictors, the second PC explains the maximum variance of all $P$ residuals, and so on. 
This means that the explained variance decreases as fast as the data allows as more PCs are retained. 
As a result, $Q$ is usually taken to be much smaller than $P$; if $Q = P$, the variance in the $P$ predictors would be redistributed across $P$ new predictors and this would defy the goal of PCA to reduce the number $P$ considerably while retaining as much variance as possible.
In practice, when estimating PCR, researchers rely on cross-validation \citep[e.g., ][]{vervloetEtAl:2016} to guide this decision.

\subsection{Multiple imputation with principal component regression}

\cite{costantiniEtAl:2023} found that the best way to incorporate PCR into MICE is to extract PCs at every iteration.
When imputing $\mathbf{z}_j$ in the $i$th iteration of MICE, the PCs can be estimated from $\mathbf{Z}^{(i)}_{-j}$ and used as predictors in the univariate imputation model.
Each univariate imputation model can then be defined as:
\begin{equation} \label{eq:MIPCR}
    f(\mathbf{z}_j|\mathbf{T}^{(i)}_{-j}, \mathbf{\theta}_j),
\end{equation}
where $\mathbf{T}^{(i)}_{-j}$ is the matrix storing the PC scores estimated on $\mathbf{Z}^{(i)}_{-j}$.
The steps described in Algorithm~\ref{alg:pcr} are followed to impute $\mathbf{z}_j$ with PCR at every iteration.
We refer to this use of PCR within MICE as MI-PCR.
This MI-PCR incorporates uncertainty around the imputation model parameters using bootstrapping following the same principle as the `imputation under the normal linear model with bootstrap' algorithm described by~\citet[][p. 69]{vanBuuren:2018}.
\begin{algorithm}
    \caption{Imputation under the PCR model with bootstrap}\label{alg:pcr}
    \begin{algorithmic}[1]
        \State For a given $\mathbf{z}_j$ variable under imputation, draw a bootstrap version $\mathbf{z}^{*}_{j, \text{obs}}$ with replacement from the observed cases $\mathbf{z}_{j, \text{obs}}$, and store as $\mathbf{Z}^{*}_{-j, \text{obs}}$ the corresponding rows of $\mathbf{Z}^{(i)}_{-j}$.
        \State Center and scale $\mathbf{Z}^{*}_{-j, \text{obs}}$ and store the result as $\tilde{\mathbf{Z}}^{*}_{-j, \text{obs}}$
        \State Center and scale $\mathbf{Z}^{(i)}_{-j, \text{mis}}$ based on the means and standard deviations of $\mathbf{Z}^{*}_{-j, \text{obs}}$ and store the result as $\tilde{\mathbf{Z}}_{-j, \text{mis}}$
        \State Center $\mathbf{z}^{*}_{j, \text{obs}}$ on its mean value $\bar{\mathbf{z}}^{*}_{j, \text{obs}}$ and store it in $\tilde{\mathbf{z}}^{*}_{j, \text{obs}}$.
        \State Estimate $\mathbf{W}$ and $\mathbf{P}$ by the eigendecomposition of the cross-product matrix of $\tilde{\mathbf{Z}}^{*}_{-j, \text{obs}}$
        \State Compute the first $Q$ PCs as $\mathbf{T}^{(i)}_{-j, \text{obs}} = \tilde{\mathbf{Z}}^{*}_{-j, \text{obs}} \mathbf{W}$
        \State Regress the mean-centered $\tilde{\mathbf{z}}^{*}_{j, \text{obs}}$ on $\mathbf{T}^{(i)}_{-j, \text{obs}}$ and store the regression coefficients $\mathbf{\beta}$
        \State Estimate the residual error variance $\sigma^2$ as the ratio between the residual sum of square ($RSS$) and the degrees of freedom ($df$):
        \begin{align*}
            \sigma^2 & = RSS / df                                                                   \\
                     & = \frac{\Sigma \, (\tilde{\mathbf{z}}^{*}_{j, \text{obs}} - \mathbf{T}^{(i)}_{-j, \text{obs}} \mathbf{\beta})^2}{N - Q}
        \end{align*}
        \State Obtain the predicted values for $\mathbf{z}_{j, miss}$ by
        \begin{equation*}
            \hat{z}_{j, miss} = \tilde{\mathbf{Z}}_{-j, \text{mis}} \mathbf{W} \mathbf{\beta}
        \end{equation*}
        \State Obtain imputations by adding normally distributed errors scaled by $\sigma^2$ to these predictions and by adding $\bar{\mathbf{z}}^{*}_{j, \text{obs}}$ to center them appropriately.
    \end{algorithmic}
\end{algorithm}

MI-PCR allows the researcher imputating the data to include all predictors in the imputation model for every variable, bypassing the difficult model selection step, while preserving the advantages of an inclusive strategy.
However, the performance of MI-PCR is highly sensitive to the number of PCs computed.
Not using enough PCs to adequately represent the latent structure (i.e., using fewer PCs than the true number of latent variables) will produce poor imputations~\citep{costantiniEtAl:2023}.
Furthermore, there is no guarantee the number of PCs that optimally represent $\mathbf{Z}^{(i)}_{-j}$ will be good imputation model predictors as the PCs retained might be summarizing information that is unrelated to the variable under imputation $\mathbf{z}_j$.
Finally, performing PCA on large datasets involves demanding matrix operations.
MI-PCR requires repeating these intensive manipulations for every variable under imputation and every iteration of the MICE algorithm.

\subsection{Multiple imputation with supervised dimensionality reduction}

SDR techniques represent an alternative to PCA that could obviate some of the limitations of MI-PCR outlined above.
In particular, the \emph{superivsion} of SDR methods should help computing PCs that are better predictors of the variables under imputation than the ones produced by PCA, and, as a corollary, it could also allow to retain fewer PCs than the number of latent variables in the data-generating model.
In what follows we describe three alternative approaches to finding linear combinations of the predictors that do a good job of both summarizing the predictors and predicting the outcome variable.
For each approach, we first describe how it works, and then we describe its implementation as a univariate imputation method in the MICE algorithm.

\subsubsection{Supervised principal component regression}\label{subsec:meth-spcr}

\cite{bairEtAl:2006} proposed computing the PCs only on the subset of variables that are associated with the dependent variable.
Their approach is straightforward:
\begin{enumerate}
    \item Regress $\mathbf{y}$ onto each column of $\mathbf{X}$ via $P$ separate simple linear regressions. 
    Because the data are standardized, the regression coefficients of these simple linear regression are equivalent to correlation coefficients. 
    The strength of the association is what matters in the predictive task, so we consider only the absolute value of the correlation and refer to it as $\hat{\rho}$.
    \item Define the subset $\mathbf{X}_{s} \in \mathbf{X}$ by discarding all variables whose correlation $\hat{\rho}$ is less than a selected threshold $\rho_s$.
    \item Use $\mathbf{X}_{s}$ to estimate the PCs.
    \item Use these PCs as independent variables in the PCR model.
\end{enumerate}
A key aspect of the method is that both the number of PCs and the threshold value $\rho_s$ can be determined by cross-validation.
We refer to this approach as supervised principal component regression (SPCR).

In SPCR, the component weights are estimated by minimizing the same criterion as in Equation~\ref{eq:pca}, but only a subset of relevant variables from $\mathbf{X}$ is used for the computation.
By doing so, SPCR effectively sets to 0 component weights for variables that are not relevant predictors of $\mathbf{y}$.
As a result, SPCR produces PCs that are better predictors of $\mathbf{y}$ and improves the predictive performance of PCR.
We refer to the approach of excluding variables that are uninteresting for the prediction of the dependent variable as \emph{discrete} supervision.

A similar approach, known as Sparse PCA~\citep{zouEtAL:2006}, reduces the number of variables explicitly used in the PC computation by combining a lasso penalty with the PCA optimization criterion.
Similarly to SPCR, Sparse PCA sets certain loadings to 0, but it does so to increase the interpretability of the resulting PCs, not to improve their predictive performance.
This key difference makes SPCR a more suitable tool than Sparse PCA for aiding automatic imputation model specification.
Therefore, we considered SPCR and not Sparse PCA as a means to reduce the dimensionality of the imputation models.

In the context of imputation, SPCR can be used as a univariate imputation model in a similar way to PCR.
For each partially observed $\mathbf{z}_j$, with $j \in \{1, \dots, J\}$, the imputation model can be defined as:
\begin{equation} \label{eq:uim-spcr}
    f(\mathbf{z}_j|\mathbf{T}^{(i)}_{s}, \mathbf{\theta}_j),
\end{equation}
where $\mathbf{T}^{(i)}_{s}$ is the matrix of PCs computed on $\mathbf{Z}^{(i)}_{s}$, the subset of variables with $\hat{\rho} > \rho_s$, at the $i$th iteration of the MICE algorithm.
The steps described in Algorithm~\ref{alg:mispcr} are followed to impute $\mathbf{z}_j$ at every iteration.
We refer to this use of SPCR within MICE as MI-SPCR.
\begin{algorithm}
    \caption{Imputation under the SPCR model with bootstrap}\label{alg:mispcr}
    \begin{algorithmic}[1]
        \State For a given $\mathbf{z}_j$ variable under imputation, draw a bootstrap version $\mathbf{z}^{*}_{j, \text{obs}}$ with replacement from the observed cases $\mathbf{z}_{j, \text{obs}}$, and store as $\mathbf{Z}^{*}_{-j, \text{obs}}$ the corresponding values on $\mathbf{Z}^{(i)}_{-j, \text{obs}}$.
        \State Compute the absolute correlation between $\mathbf{z}^{*}_{j, \text{obs}}$ and every potential predictor in $\mathbf{Z}^{*}_{-j, \text{obs}}$.
        \State For every $\rho_h$, create a set of predictors with absolute correlation higher than $\rho_h$.
        \State Use \emph{K}-fold cross-validation to select the value of $\rho_h$ and the associated set of predictors that return the PCR model with the smallest prediction error. 
        Define $\rho_s$ as the selected value.
        \State Drop from $\mathbf{Z}^{*}_{-j, \text{obs}}$ and $\mathbf{Z}^{(i)}_{-j, \text{mis}}$ all variables with an absolute correlation smaller than $\rho_s$ and create $\mathbf{Z}^{*}_{s, obs}$ and $\mathbf{Z}_{s, mis}$.
        \State Center and scale $\mathbf{Z}^{*}_{s, obs}$ and store the result as $\tilde{\mathbf{Z}}^{*}_{s, obs}$
        \State Center and scale $\mathbf{Z}_{s, mis}$ based on based on the means and standard deviations of $\mathbf{Z}^{*}_{s, obs}$ and store the result as $\tilde{\mathbf{Z}}_{s, mis}$
        \State Center $\mathbf{z}^{*}_{j, \text{obs}}$ on its mean value $\bar{\mathbf{z}}^{*}_{j, \text{obs}}$ and store it in $\tilde{\mathbf{z}}^{*}_{j, \text{obs}}$.
        \State Estimate $\mathbf{W}$ and $\mathbf{P}$ by the eigendecomposition of the cross-product matrix of $\tilde{\mathbf{Z}}^{*}_{s, obs}$
        \State Compute the first $Q$ PCs as $\mathbf{T}^{(i)}_{s, obs} = \tilde{\mathbf{Z}}^{*}_{s, obs} \mathbf{W}$
        \State Regress the mean centered $\tilde{\mathbf{z}}^{*}_{j, \text{obs}}$ on $\mathbf{T}^{(i)}_{s, obs}$ and store the regression coefficients $\mathbf{\beta}$.
        \State Estimate the residual error variance $\sigma^2$ as the ratio between the residual sum of square ($RSS$) and the degrees of freedom ($df$):
        \begin{align*}
            \sigma^2 & = RSS / df                                                                   \\
                     & = \frac{\Sigma \, (\tilde{\mathbf{z}}^{*}_{j, \text{obs}} - \mathbf{T}^{(i)}_{s, obs} \mathbf{\beta})^2}{N - Q}
        \end{align*}
        \State Obtain the predicted values for $\mathbf{z}_{j, miss}$ by
        \begin{equation*}
            \hat{z}_{j, miss} = \tilde{\mathbf{Z}}_{s, mis} \mathbf{W} \mathbf{\beta}
        \end{equation*}
        \State Obtain imputations by adding normally distributed errors scaled by $\sigma^2$ to these predictions and by adding $\bar{\mathbf{z}}^{*}_{j, \text{obs}}$ to center them appropriately.
    \end{algorithmic}
\end{algorithm}

In our implementation of MI-SPCR, \emph{K}-fold cross-validation \citep[][pp. 241--245]{hastieEtAl:2009} is used to select $\rho_s$, from a user-defined vector of possible values.
For every threshold value in the interval $[0, 1]$, all predictors of $\mathbf{z}^{*}_{j, \text{obs}}$ in $\mathbf{Z}^{*}_{-j, \text{obs}}$ with a correlation larger than the threshold form an active set of predictors.
Then, $Q$ PCs are extracted from each active set and used to predict $\mathbf{z}^{*}_{j, \text{obs}}$ in a \emph{K}-fold cross-validation procedure.
The active set giving the lowest cross-validated prediction error is kept.
As with MI-PCR, the number of components $Q$ is considered fixed, but it can be selected by the same cross-validation procedure.
Note that for a given number of components, only certain threshold values are allowed.
We can compute $Q$ components only if the data have at least $Q$ columns.
Therefore, the more components we want to estimate, the less restrictive $\rho_s$ can be.
If we ask for as many components as there are columns in the data, then $\rho_s$ must be large enough to keep all columns of the data, making MI-SPCR equivalent to MI-PCR.

\subsubsection{Principal covariates regression}\label{subsec:meth-pcovr}

Principal covariates regression \citep[PCovR, ][]{deJongKiers:1992} is an SDR approach that modifies the optimization criteria behind PCA to include information from the outcome variable in the optimization problem.
PCovR looks for a low-dimensional representation of $\mathbf{X}$ that accounts for the maximum amount of variation in both $\mathbf{X}$ and $\mathbf{y}$.
To understand how PCovR differs from PCR consider the following decomposition of the data:
\begin{align}
    \mathbf{X} & = \mathbf{T} \mathbf{P}'_{\mathbf{X}} + \mathbf{E}_{\mathbf{X}} \\
    \mathbf{y} & = \mathbf{T} \mathbf{P}'_{\mathbf{y}} + \mathbf{e}_{\mathbf{y}} \\
    \mathbf{T} & = \mathbf{X} \mathbf{W}               
\end{align}
where $\mathbf{T}$ and $\mathbf{W}$ are defined as in~\ref{eq:pca}, $\mathbf{P}_{\mathbf{X}}$ is $\mathbf{P}$ from~\ref{eq:pca}, and $\mathbf{P}_{\mathbf{y}}$ is the $Q \times 1$ vector of weights relating $\mathbf{y}$ to the component scores in $\mathbf{T}$.
$\mathbf{E}_{\mathbf{X}}$ and $\mathbf{e}_{\mathbf{y}}$ are reconstruction errors.
They represent the information lost by using $\mathbf{T}$ as a summary of $\mathbf{X}$ and the errors in the linear regression model, respectively.
PCovR can be formulated as the task of minimizing a weighted combination of both $\mathbf{E}_{\mathbf{X}}$ and $\mathbf{e}_{\mathbf{y}}$:
\begin{equation}\label{eq:PCovR}
    (\mathbf{W}, \mathbf{P}_{\mathbf{X}}, \mathbf{P}_{\mathbf{y}}) = \underset{\mathbf{W}, \mathbf{P}_{\mathbf{X}}, \mathbf{P}_{\mathbf{y}}}{\operatorname{argmin}} \; \alpha \, \lVert (\mathbf{X} - \mathbf{XWP}_{\mathbf{X}}') \rVert^2 + (1 - \alpha) \, \lVert (y - \mathbf{XWP}_{\mathbf{y}}') \rVert^2
\end{equation}
subject to the constraint $\mathbf{W}'\mathbf{X}' \mathbf{XW} = \mathbf{T}' \mathbf{T} = \mathbf{I}$.

The $\alpha$ parameter defines which reconstruction error is being prioritized.
When $\alpha = 1$, the emphasis is exclusively placed on reconstructing $\mathbf{X}$, casting PCR as a special case of PCovR.
When $\alpha = 0.5$, the importance of $\mathbf{X}$ and $\mathbf{y}$ is equally weighted, a case that resembles partial least square regression (PLSR), which we discuss in subsection~\ref{subsec:pls}.
In practice, the value of $\alpha$ can be found by cross-validation or according to a sequential procedure based on maximum likelihood principles \citep{vervloetEtAl:2013}.
In particular, 
\begin{equation}\label{eq:aml}
    \alpha_{ML} = \frac{\lVert \mathbf{X} \lVert^2}{\lVert \mathbf{X} \lVert^2  + \lVert y \lVert^2 \frac{\hat{\sigma}_{\mathbf{E}_{\mathbf{X}}}^2}{\hat{\sigma}_{e_{\mathbf{y}}}^2}}
\end{equation}
where $\hat{\sigma}_{\mathbf{E}_{\mathbf{X}}}^2$ can be obtained as the unexplained variance by components computed according to classical PCA on $\mathbf{X}$ and $\hat{\sigma}_{e_{\mathbf{y}}}^2$ can be estimated as the unexplained variance by the linear model regressing $\mathbf{y}$ on $\mathbf{X}$.
Note that, for the same data set, the more components are retained, the smaller $\hat{\sigma}_{\mathbf{E}_{\mathbf{X}}}$ is, the higher $\alpha_{ML}$ is, and the closer PCovR becomes to PCR.
Retaining the same number of components as the number of variables in $\mathbf{X}$ results in $\hat{\sigma}_{\mathbf{E}_{\mathbf{X}}} = 0$ and $\alpha = 1$, which casts PCR as a special case of PCovR.

Compared to PCR, PCovR allows estimating PCs that not only represent well the predictor variables but also predict well the dependent variable.
Compared to SPCR, the PCs computed with PCovR are always linear combinations of \emph{all} variables in $\mathbf{X}$.
PCovR can downweigh irrelevant variables for the prediction of $\mathbf{y}$, but it will never exclude them entirely.
We refer to the PCovR approach to supervision as \emph{continuous}, as opposed to the \emph{discrete} supervision of SPCR.

When applied to the MICE algorithm, we can use PCovR as a univariate imputation model in a similar way to how we can use PCR and SPCR.
At every iteration of the MICE algorithm, the steps described in Algorithm~\ref{alg:mipcovr} are followed to impute $\mathbf{z}_j$.
We refer to this use of PCovR within MICE as MI-PCovR.
\begin{algorithm}
    \caption{Imputation under the PCovR model with bootstrap}\label{alg:mipcovr}
    \begin{algorithmic}[1]
        \State For a given $\mathbf{z}_j$ variable under imputation, draw a bootstrap version $\mathbf{z}^{*}_{j, \text{obs}}$ with replacement from the observed cases $\mathbf{z}_{j, \text{obs}}$, and store as $\mathbf{Z}^{*}_{-j, \text{obs}}$ the corresponding values on $\mathbf{Z}^{(i)}_{-j}$.
        \State Center and scale $\mathbf{Z}^{*}_{-j, \text{obs}}$ and store the result as $\tilde{\mathbf{Z}}^{*}_{-j, \text{obs}}$.
        \State Center and scale $\mathbf{Z}^{(i)}_{-j, \text{mis}}$ based on based on the means and standard deviations of $\mathbf{Z}^{*}_{-j, \text{obs}}$ and store the result as $\tilde{\mathbf{Z}}_{-j, \text{mis}}$.
        \State Center $\mathbf{z}^{*}_{j, \text{obs}}$ on its mean value $\bar{\mathbf{z}}^{*}_{j, \text{obs}}$ and store it in $\tilde{\mathbf{z}}^{*}_{j, \text{obs}}$.
        \State Compute the value of $\alpha$ based on \ref{eq:aml}.
        \State Compute $Q$ PCs by estimating the PCovR $\mathbf{W}$, $\mathbf{P}_{-j}$, and $\mathbf{P}_{j}$ based on $\tilde{\mathbf{z}}^{*}_{j, \text{obs}}$ and $\tilde{\mathbf{Z}}^{*}_{-j, \text{obs}}$. Note that $\mathbf{P}_{-j}$, and $\mathbf{P}_{j}$ correspond to $\mathbf{P}_{\mathbf{X}}$, and $\mathbf{P}_{\mathbf{y}}$ of equation \eqref{eq:PCovR}.
        \State Estimate the residual error variance $\sigma^2$ as the ratio between the residual sum of square ($RSS$) and the degrees of freedom ($df$):
        \begin{align*}
            \sigma^2 & = RSS / df                                                                   \\
                     & = \frac{\Sigma \, (\tilde{\mathbf{z}}^{*}_{j, \text{obs}} - \tilde{\mathbf{Z}}^{*}_{-j, \text{obs}} \mathbf{W} \mathbf{P}_{j}')^2}{N - Q}
        \end{align*}
        \State Obtain the predicted values for $\mathbf{z}_{j, miss}$ by
        \begin{equation*}
            \hat{z}_{j, miss} = \tilde{\mathbf{Z}}_{-j, \text{mis}} \mathbf{W} \mathbf{P}_{-j}'
        \end{equation*}
        \State Obtain imputations by adding normally distributed errors scaled by $\sigma^2$ to these predictions and by adding $\bar{\mathbf{z}}^{*}_{j, \text{obs}}$ to center them appropriately.
    \end{algorithmic}
\end{algorithm}

\subsubsection{Partial least square regression}\label{subsec:pls}

Partial least square regression \citep[PLS,][]{wold:1975} is a dimensionality reduction technique that seeks linear combinations (or PLS components) that account for a large proportion of the variance in the predictors and correlate strongly with the dependent variable.
Like PCR, PLSR finds independent linear combinations of the predictors in $\mathbf{X}$ that summarize the data well and uses these linear combinations to predict the dependent variable.
Like PCovR, the weights defining the linear combinations of the predictors are computed using all the predictor variables and the outcome, resulting in a \emph{continuous} supervision: irrelevant variables will be downweighted in the linear combinations, but they will not be completely ignored.
Unlike PCR, SPCR, and PCovR, PLSR computes one linear combination at a time and stops at the required number of PLS components $Q$.

PLSR estimates $\mathbf{t}_{1}$, the first PLS component, by:
\begin{enumerate}
    \item Computing the vector of weights $\mathbf{w}_1$ with elements $w_{1, j} = \mathbf{x}_j' \mathbf{y}$ for $j \in \{1, \dots, P\}$, the inner products between each predictor $\mathbf{x}_j$ and the dependent variable $\mathbf{y}$.
    \item Deriving the constructed variable $\mathbf{t}_1 = \sum_{j=1}^{P}\mathbf{x}_j w_{1, j} = \mathbf{X} \mathbf{w}_1$.
    \item Orthogonalizing $\mathbf{x}_1$ to $\mathbf{x}_P$ with respect to $\mathbf{t}_1$.
\end{enumerate}
The second linear combination ($\mathbf{t}_2 = \mathbf{X} \mathbf{w}_2$) is then derived by repeating the same procedure but replacing each $\mathbf{x}_j$ with their versions orthogonalized with respect to $\mathbf{t}_1$.
In PLSR, the $q$th weight vector ($\mathbf{w}_q$) maximizes the following optimization criterion~\citep{stoneBrooks:1990, frankFriedman:1993}:
\begin{equation}
    \underset{\mathbf{w}_{q}}{\operatorname{argmax}} \; \text{Corr}^2(\mathbf{y}, \mathbf{X}\mathbf{w}_{q}) \text{Var}(\mathbf{X}\mathbf{w}_{q})
\end{equation}
where $\text{Corr}^2(.)$ is the squared correlation of the vectors between brackets.
As with all other methods, the linear combinations derived by the PLS algorithm are constrained to be mutually orthogonal.

As with SPCR and PCovR, at every iteration of the MICE algorithm, we can use PLSR to obtain imputations.
Algorithm~\ref{alg:miplsr} describes the univariate imputation method based on PLSR\footnote{A similar version of PLS as a univariate imputation method has also been implemented in the R package `miceadds' \citep{miceadds}.} that we used to impute $\mathbf{z}_j$.
We refer to this use of PLSR within MICE as MI-PLSR.
In Table~\ref{tab:meth-sum}, we summarize the differences between the univariate imputation methods used by MI-PCR and the SDR-based approaches we described.
\begin{algorithm}
    \caption{Imputation under the PLSR model with bootstrap}\label{alg:miplsr}
    \begin{algorithmic}[1]
        \State For a given $\mathbf{z}_j$ variable under imputation, draw a bootstrap version $\mathbf{z}^{*}_{j, \text{obs}}$ with replacement from the observed cases $\mathbf{z}_{j, \text{obs}}$, and store as $\mathbf{Z}^{*}_{-j, \text{obs}}$ the corresponding values on $\mathbf{Z}^{(i)}_{-j}$.
        \State Estimate PLSR with $Q$ components by regressing $\mathbf{z}^{*}_{j, \text{obs}}$ onto $\mathbf{Z}^{*}_{-j, \text{obs}}$.
        \State Estimate the residual error variance $\sigma^2$ as the ratio between the residual sum of square ($RSS$) and the degrees of freedom ($df$).
        \State Obtain the predicted values for $\mathbf{z}_{j, miss}$ based on the trained PLSR model.
        \State Obtain imputations by adding noise scaled by $\sigma^2$ to these predictions.
    \end{algorithmic}
\end{algorithm}
\begin{table}
    \newcommand{\criteriaPCR}{\underset{\mathbf{W}, \mathbf{P}}{\operatorname{argmin}} \; \lVert \tilde{\mathbf{Z}}^{*}_{-j, \text{obs}} - \tilde{\mathbf{Z}}^{*}_{-j, \text{obs}}\mathbf{WP}' \rVert^2}
    \newcommand{\criteriaSPCR}{\underset{\mathbf{W}, \mathbf{P}}{\operatorname{argmin}} \; \lVert \tilde{\mathbf{Z}}^{*}_{s, obs} - \tilde{\mathbf{Z}}^{*}_{s, obs}\mathbf{WP}' \rVert^2}
    \newcommand{\criteriaPCovR}{$\begin{aligned} \underset{\mathbf{W}, \mathbf{P}_{\mathbf{X}}, \mathbf{P}_{\mathbf{y}}}{\operatorname{argmin}} & \; \alpha \, \lVert (\tilde{\mathbf{Z}}^{*}_{-j, \text{obs}} - \tilde{\mathbf{Z}}^{*}_{-j, \text{obs}}\mathbf{WP}_{\mathbf{X}}') \rVert^2 + \\ & + (1 - \alpha) \, \lVert (\tilde{\mathbf{z}}^{*}_{j, \text{obs}} - \tilde{\mathbf{Z}}^{*}_{-j, \text{obs}}\mathbf{WP}_{\mathbf{y}}') \rVert^2 \end{aligned}$}
    \newcommand{\criteriaPLSR}{\underset{\mathbf{w}_{q}}{\operatorname{argmax}} \; \text{Corr}^2(\tilde{\mathbf{z}}^{*}_{j, \text{obs}}, \tilde{\mathbf{Z}}^{*}_{-j, \text{obs}}\mathbf{w}_{q}) \text{Var}(\tilde{\mathbf{Z}}^{*}_{-j, \text{obs}}\mathbf{w}_{q})}
    \centering
    \resizebox{\textwidth}{!}{
        \begin{tabular}{c c c c c}
            \toprule
            Method   & \begin{tabular}{cc}Supervision\\type\end{tabular} & \begin{tabular}{cc}Optimization\\criterion\end{tabular}        & \begin{tabular}{cc}Estimated\\parameters\end{tabular} & \begin{tabular}{cc}Tuning\\parameters\end{tabular}               \\
            \midrule
            MI-PCR   & none             & \(\displaystyle \criteriaPCR \)   & \(\displaystyle \mathbf{W}, \mathbf{P} \)                                       & -           \\
            MI-SPCR  & discrete         & \(\displaystyle \criteriaSPCR \)  & \(\displaystyle \mathbf{W}, \mathbf{P} \)                                       & \(\displaystyle  \rho_s   \) \\
            MI-PCovR & continuous       & \criteriaPCovR & \(\displaystyle \mathbf{W}, \mathbf{P}_{\mathbf{X}}, \mathbf{P}_{\mathbf{y}} \) & \(\displaystyle \alpha   \) \\
            MI-PLSR  & continuous       & \(\displaystyle \criteriaPLSR \)  & \(\displaystyle \mathbf{w}_q \)                                                 & -           \\
            \bottomrule
        \end{tabular}
    }
    \caption{
        Summary of the differences between the univariate imputation models used within the MICE algorithm.
    }
    \label{tab:meth-sum}
\end{table}

\section{Simulation study}\label{sec:sim-study}

We investigated the relative performance of unsupervised PCR and the supervised alternatives described above with a Monte Carlo simulation study.
In particular, we investigated the estimation bias, confidence interval width, and confidence interval coverage of several analysis model parameters obtained after imputation.
We varied the dimensionality of the latent structure in the data-generating model, the proportion of missing values, the missing data mechanism, and the number of components used as predictors in the imputation models.
The proportion of missing values and the missing data mechanism both influence the statistical properties of any missing data treatment, while the number of latent variables in the data-generating model and the number of components used in the imputation model affect the extent to which supervision can improve upon the limitations of MI-PCR.
In table~\ref{tab:cond}, we summarize the experimental factors we varied and the levels we used with each factor.

\begin{table}
  \centering
  \begin{tabular}{c c p{5cm}}
    \toprule
    Experimental factor & Label & Levels \\
    \midrule
    Number of latent variables      & $L$   & 2, 10, 50 \\
    Missing data mechanism used  & $mech$  & MCAR, MAR \\
    Proportion of missing values    & $pm$ & 0.1, 0.25, 0 50 \\
    Missing data treatment &  $method$      & MI-PCR, MI-SPCR, MI-PCovR, MI-PLS, MI-QP, MI-AM, MI-ALL, CC, FO \\
    Number of components    &   $nc$        & 0, 1 to 12, 20, 29, 30, 40, 48, 49, 50, 51, 52, 60, 149         \\
    \bottomrule
  \end{tabular}
  \caption{
    Summary of experimental factors for the simulation study.
  }
  \label{tab:cond}
\end{table}

\subsection{Procedure}

The simulation study involved four steps:
\begin{enumerate}
  \item Data generation: We generated $R = 240$ data sets from a confirmatory factor analysis model, following the procedure described in Section \ref{subsubsec:data-generation}.
  \item Missing data generation: We generated missing values on three target items in each data set, following the procedure described in Section~\ref{subsubsec:missing-data-generation}.
  \item Imputation: We generated $d = 5$ multiply imputed versions of each generated data set using different imputation methods, as described in Section~\ref{subsubsec:imputation-procedures}.
  \item Analysis: We estimated the means, variances, covariances, and correlations of the three items with missing values on the $d$ imputed data sets, and we pooled the estimates according to Rubin's rules \citep[p. 76]{rubin:1986}.
        We then assessed each imputation method by computing the bias of different parameter estimates, and their confidence interval widths and coverages as described in Section~\ref{subsubsec:analysis-and-outcome-measures}.
\end{enumerate}

\subsubsection{Data generation}\label{subsubsec:data-generation}

For each of the $R$ replications, we generated a $1000 \times P$ data matrix $\mathbf{Z}$.
The sample size should be large enough to generate data sets that have statistical properties similar to large social science data sets.
Each data set was generated based on the following model:
\begin{equation} \label{eq:CFAmod}
  \mathbf{Z} = \mathbf{F}\mathbf{\Lambda}' + \mathbf{E},
\end{equation}
where $\mathbf{F}$ is a $1000 \times L$ matrix of latent variables scores, $L$ is the number of latent variables, $\mathbf{\Lambda}$ is a $3 \times L$ matrix of factor loadings, where $3$ is the number of items measuring each latent variable, and $\mathbf{E}$ is a $1000 \times P$ matrix of measurement errors, where $P = 3 * L$.
The dimensionality of the data resembles that of the many large social surveys that use short scales to measure respondents' attitudes such as \emph{political engagement} and \emph{anti-immigrant attitudes}.
For example, consider the European Values Study~\citep{EVS:2017} which measures a variety of attitudes with 3, 4, or 5 items.

The factor loading matrix $\mathbf{\Lambda}$ in Equation~\ref{eq:CFAmod} described a simple structure~\citep[p. 234]{bollen:1989} where each item loads on exactly one of the $L$ latent variables.
In real data applications, this factor structure is uncommon but not implausible.
For example, a relatively clear simple structure can be found when analyzing personality inventories \citep[e.g., NEO-PR-I,][]{costaEtAl:1991} with the \emph{Neuroticism-Extroversion-Openness} three-factor model~\citep{mcCraeCosta:1983}.
A simple structure is also often assumed when performing exploratory factor analysis because it provides the most parsimonious explanation~\citep[pp. 183-184]{costaMcCrae:2008}.

We sampled $\mathbf{F}$ from a multivariate normal distribution with mean $\mathbf{0}$ and covariance matrix $\mathbf{\Psi}$.
The correlation between the first and second latent variables was fixed at $0.8$, while the correlation between all other latent variables and the first two was fixed at $0.1$.
Together with factor loadings fixed at $\lambda = 0.85$, these choices resulted in correlations of approximately $0.72$, $0.58$, and $0.07$ between items measuring the same latent variable, items measuring the first and second latent variable, and items measuring the first latent variable and the others, respectively.
These values represent plausible, but reasonably high, item-scale associations and they should mitigate the impact of measurement error on our findings without resorting to implausibly precise data.

The matrix of measurement errors $\mathbf{E}$ was sampled from a multivariate normal distribution with mean vector $\mathbf{0}$ and covariance matrix $\mathbf{\Theta}$.
The off-diagonal elements of $\mathbf{\Theta}$ were set to 0 to reflect uncorrelated errors, while the diagonal elements were specified as $1 - \lambda^2$ to give the simulated items unit variances.
After sampling, the columns of $\mathbf{Z}$ were rescaled to have approximately a mean of 5 and a variance of 6.5, which are common values for Likert items in social surveys measured on a 10-point scale~\cite[for example in][]{EVS:2017}.

In this data-generating procedure, we considered the number of latent variables used $(L)$ as an experimental factor with levels $2, 10, 50$, resulting in data sets containing 6, 30, and 150 total items.
\cite{costantiniEtAl:2023} showed that the number of components used in MI-PCR needs to be at least as high as the number of latent variables in the data-generating model. 
So, we expected MI-PCR to require at least $L$ components to achieve satisfactory performance. 
One of this study's main objectives was to understand how well supervision can overcome this limitation of MI-PCR.
We generated data according to a confirmatory factor analysis model instead of generating the data directly based on \text{true} principal components to avoid the misleading results that can occur when using a single model for both data generation and imputation \citep[see ][p. 4]{obermanVink:2023}.

\subsubsection{Missing data generation}\label{subsubsec:missing-data-generation}

We generated missing data on the three items measuring the first latent variable ($\mathbf{z}_{1}$, $\mathbf{z}_{2}$, $\mathbf{z}_{3}$).
The proportion of missing values per variable ($pm$) was defined as an experimental factor taking three levels $pm \in \{0.1, 0.25, 0.5\}$.
The missing data mechanism ($mech$) was a factor with two levels:
\begin{itemize}
  \item Missing completely at random (MCAR): To test how the methods performed in the simplest possible missing data mechanism, we generated
  missing values on each item based on a missing data indicator ($\delta$) sampled from a binomial distribution with success probability $pm$.
  If $\delta = 1$, the item score was set to missing. 
  If $\delta = 0$, the item score was set to observed.
        As a result, every variable had a proportion of missing values approximately equal to $pm$.
  \item Missing at random (MAR): To test how the methods performed in a more realistic situation, we generated missing values based on a MAR mechanism with the three items measuring the second latent variable ($\mathbf{z}_{4}$, $\mathbf{z}_{5}$, $\mathbf{z}_{6}$) used as predictors of missingness.
        We sampled $\delta$ from Bernoulli distributions with probabilities defined based on the following logit model:
        \begin{equation} \label{eq:logit}
          logit(\delta = 1) = \beta_0 + \mathbf{Z}_{(4, 5, 6)}\mathbf{\beta},
        \end{equation}
        where $\beta_0$ is an intercept parameter, and $\mathbf{\beta}$ is a vector of slope parameters.
        All slopes in $\mathbf{\beta}$ were fixed to 1, while the value of $\beta_0$ was chosen with an optimization algorithm that minimized the difference between the actual and desired proportion of missing values on the variable. 
        The pseudo R-squared for the logistic regression of the missing value indicator on the predictors of missingness was approximately 14\%. The AUC for the logistic regression was approximately 0.74.
        To create realistic missing data patterns, the location of missing data was fixed to right for $\mathbf{z}_1$, left for $\mathbf{z}_2$, and tails for $\mathbf{z}_3$.

\end{itemize}

\subsubsection{Imputation}\label{subsubsec:imputation-procedures}

We imputed the missing values using the four dimension reduction-based methods described above (MI-PCR, MI-SPCR, MI-PCovR, MI-PLSR) as well as three traditional approaches:
\begin{itemize}
  \item MI with all the available variables used as predictors in the imputation models (MI-ALL)\@, which represents the most naive way to define the imputation model.
  \item MI with a correlation-based threshold strategy to select the subset of important predictors (MI-QP)\@.
        As a pragmatic point of comparison, this method used the \emph{quickpred} function from the R package mice \citep{mice} to select the predictors for the univariate imputation models via the correlation-based threshold strategy described by \citet[pp. 687-–688]{vanBuurenEtAl:1999}.
        To implement this approach, we selected only those predictors that correlated with the imputation targets (or their associated missingness indicators) higher than $0.1$.
  \item MI with the analysis model variables used as sole predictors in the imputation models (MI-AM)\@.
        This method produces the simplest possible congenial imputation model and is interesting due to its popularity in the social scientific literature~\citep{costantiniEtAl:2023b}.
\end{itemize}
As reference points, we also treated the missing values with complete case analysis (CC) and estimated the analysis model from the original, fully observed data.

Every imputation-based method used simple random draws from the observed data as starting values and was run to obtain 5 imputed data sets.
Convergence was achieved after 20 iterations for all methods\footnote{Convergence plots are reported in the interactive results dashboard that we developed to accompany this article. See Section~\ref{subsec:results} for more details.}.
All our dimensionality reduction-based algorithms need the user to define $Q$, the optimal number of components.
In practice, researchers will choose a single value of $Q$ with a cross-validation procedure, but in this study, we are interested in evaluating the performance of the four dimension-reduction imputation approaches while varying the number of retained components.
Therefore, we defined $Q$ as one of our main experimental factors (the number of components, $nc$) taking values $\{1, \ldots, 12, 20, 29, 30, 40, 48, 49, 50, 51, 52, 60, 149\}$.
These values were chosen to both cover the range of possible choices (i.e., $1, \dots, [P - 1]$) and to provide more granularity around the true number of latent variables (i.e., 3, 10, 50).

Finally, for MI-SPCR, we used the cross-validation procedure described in Section~\ref{subsec:meth-spcr} to select $\rho_s$ from the vector of values $\{0.05, 0.1, 0.15, \ldots, 0.95\}$.
For a given number of components, some threshold values can exclude enough predictors to preclude computing the required number of components.
To avoid this possibility, the cross-validation algorithm only considered values of $\rho$ that retained enough variables to compute the required number of components.
The weighting parameter ($\alpha$) for MI-PCovR was selected using the sequential MLE-based estimation procedure described in~\cite{vervloetEtAl:2013}.
The degrees of freedom for the PLSR imputation model were computed based on the naive approach described by~\cite{kramerMasashi:2011}

\subsubsection{Analysis and comparison criteria}\label{subsubsec:analysis-and-outcome-measures}

For a given parameter $\phi$ (e.g., the mean of $\mathbf{z}_1$, the correlation between $\mathbf{z}_1$ and $\mathbf{z}_2$), we used the absolute percent relative bias (PRB) to quantify the estimation bias introduced by the imputation procedure:
\begin{equation} \label{eq:prb}
  \text{PRB} = \displaystyle\left\lvert\ \frac{\bar{\hat{\phi}} - \phi}{\phi} \right\rvert \times 100
\end{equation}
where $\phi$ is the true value of the focal parameter defined as
$\sum_{r=1}^{R} \hat{\phi}_{r}^{full}/R$
, with
$\phi_{r}^{full}$
being the parameter estimate for the $r$th repetition computed on the fully observed data.
The averaged focal parameter estimate under a given missing data treatment was computed as
$\bar{\hat{\phi}} = \sum_{r=1}^{R} \hat{\phi}_{r}/R$,
with
$\bar{\hat{\phi}}_{r}$ being the estimate obtained from the treated incomplete data in the
$r$th replication.
Following~\cite{muthenEtAl:1987}, we considered $\text{PRB} > 10$ as indicative of problematic
estimation bias.

To measure the statistical efficiency of the imputation methods we computed the average width of
the confidence intervals (CIW).
\begin{equation} \label{eq:ciw}
  \text{CIW} = \frac{\sum_{r=1}^{R} (\widehat{\text{CI}}^{upper}_{r} - \widehat{\text{CI}}^{lower}_{r})}{R},
\end{equation}
with $\widehat{\text{CI}}^{upper}_{r}$ and $\widehat{\text{CI}}^{lower}_{r}$ being the upper and lower bounds of the estimated confidence interval for the $r$th replication.
Narrower CIWs indicate higher efficiency.
However, narrower CIWs are not preferred if they come at the expense of good confidence interval coverage (CIC) of the parameter values.
CIC is the proportion of confidence intervals that contain the true value of the parameter, across the $R$ data samples:
\begin{equation} \label{eq:cic}
  \text{CIC} =  \frac{ \sum_{r=1}^{R} I(\phi \in \widehat{\text{CI}}_r ) }{R},
\end{equation}
where $\widehat{\text{CI}}_r$ is the confidence interval of the parameter estimate $\hat{\phi}_{r}$ in the $r$th replication, and $I(.)$ is the indicator function that returns 1 if the argument is true and 0 otherwise.
CIC depends on both the bias and the variability of the CIW for a parameter estimate.
In particular, for a given level of bias, a narrower CIW leads to lower CIC, and, for a given CIW, a larger bias leads to lower CIC.
An imputation method with good coverage should result in CICs greater than or equal to the nominal rate.
For 95\% CIs, CIC below 0.9 is usually considered problematic (e.g., \citealp[p. 52]{vanBuuren:2018}; \citealp[p. 340]{collinsEtAl:2001}) as it implies inflated Type I error rates.
High CIC (e.g., 0.99) implies inflated Type II error rates.

\subsection{Results}\label{subsec:results}
We report only the results for the correlation between $\mathbf{z}_1$ and $\mathbf{z}_2$ for the conditions with \emph{mech} = MAR and $pm = 0.5$ because the type of parameter and the different levels of these two factors did not impact the relative performances of the imputation methods.
We focused on the correlation between two items with missing values because this parameter differentiated the performances of the methods the most.
The full set of results is available via the interactive results dashboard that we developed to accompany this article \citep{costantini:2022b}.
The dashboard can be downloaded and installed as an R package, and it can be used as an R Shiny app.

\begin{figure}
  \centering
\begin{knitrout}
\definecolor{shadecolor}{rgb}{0.969, 0.969, 0.969}\color{fgcolor}

{\centering \includegraphics[width=\maxwidth]{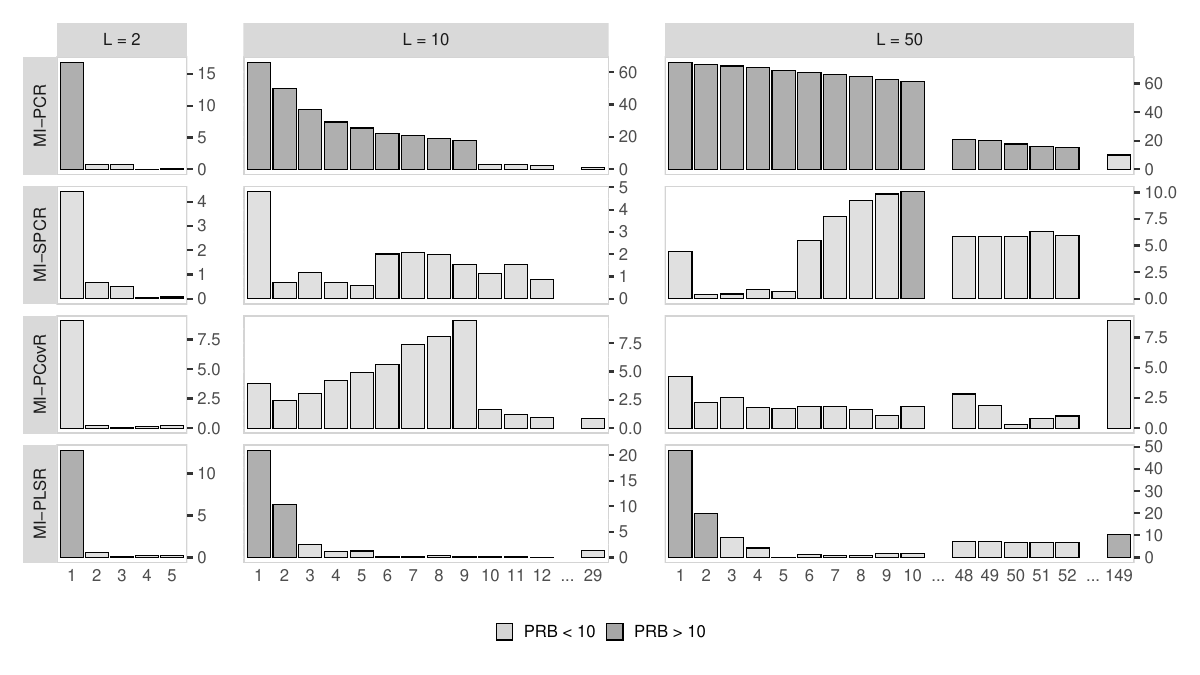} 

}

\end{knitrout}
  \caption{
    \label{fig:prb}
    The PRB for the estimated correlation coefficient between the first two items imputed is reported (Y-axis) as a function of the number of components ($nc$) used by the PCA-based imputation methods (X-axis).
    The plot is divided into a grid where the rows distinguish the results obtained after imputing the data with the four PCA-based methods and the columns distinguish the number of latent variables used to generate the data ($L$).
    All results plotted in this figure were obtained on data generated with \emph{mech} = MAR and $pm = 0.5$.
  }
\end{figure}

\begin{figure}
  \centering
\begin{knitrout}
\definecolor{shadecolor}{rgb}{0.969, 0.969, 0.969}\color{fgcolor}

{\centering \includegraphics[width=\maxwidth]{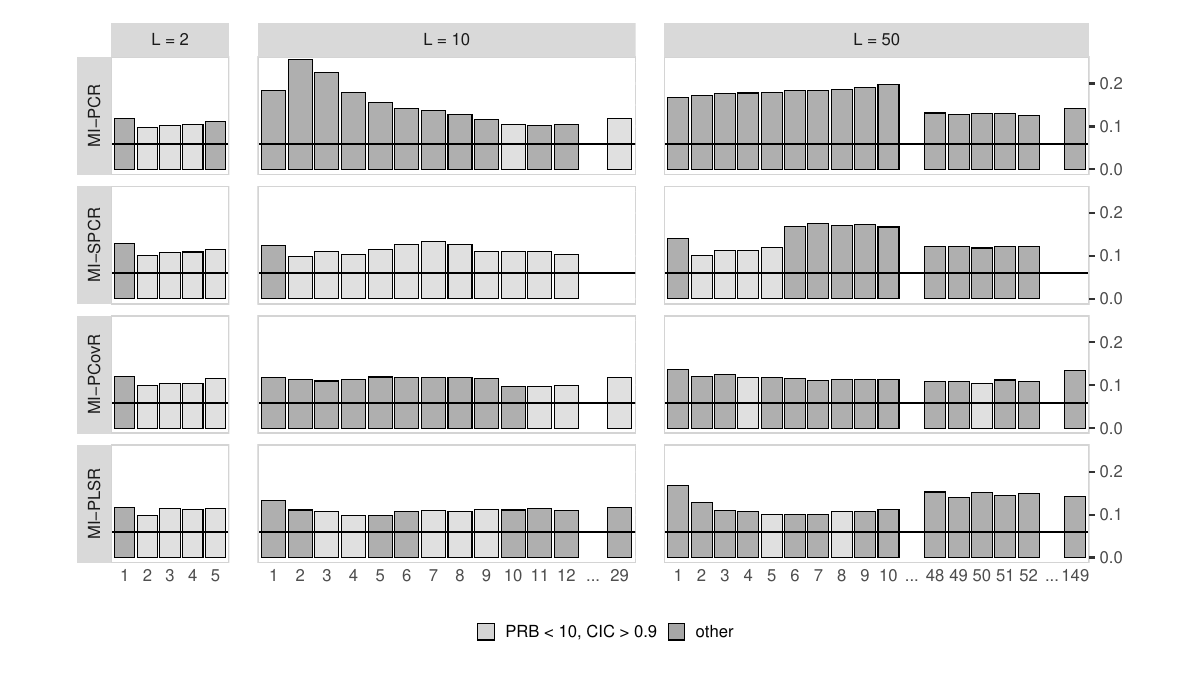} 

}

\end{knitrout}
  \caption{
    \label{fig:ciw}
    The CIW for the estimated correlation coefficient between the first two items imputed is reported (Y-axis) as a function of the number of components ($nc$) used by the PCA-based imputation methods (X-axis).
    The plot is divided into a grid where the rows distinguish the results obtained after imputing the data with the four PCA-based methods and the columns distinguish the number of latent variables used to generate the data ($L$).
    All results plotted in this figure were obtained on data generated with \emph{mech} = MAR and $pm = 0.5$.
    The light gray color indicates the parameter estimate for which the CIW is reported had both acceptable bias ($\text{PRB} < 10$) and coverage ($\text{CIC} > 0.9$).
    The dark gray color indicates the parameter estimate for which the CIW is reported had large bias ($\text{PRB} > 10$) or low coverage ($\text{CIC} < 0.9$), or both.
    The black horizontal lines represent the average CIW obtained on the original fully observed data.
  }
\end{figure}

\begin{figure}
  \centering
\begin{knitrout}
\definecolor{shadecolor}{rgb}{0.969, 0.969, 0.969}\color{fgcolor}

{\centering \includegraphics[width=\maxwidth]{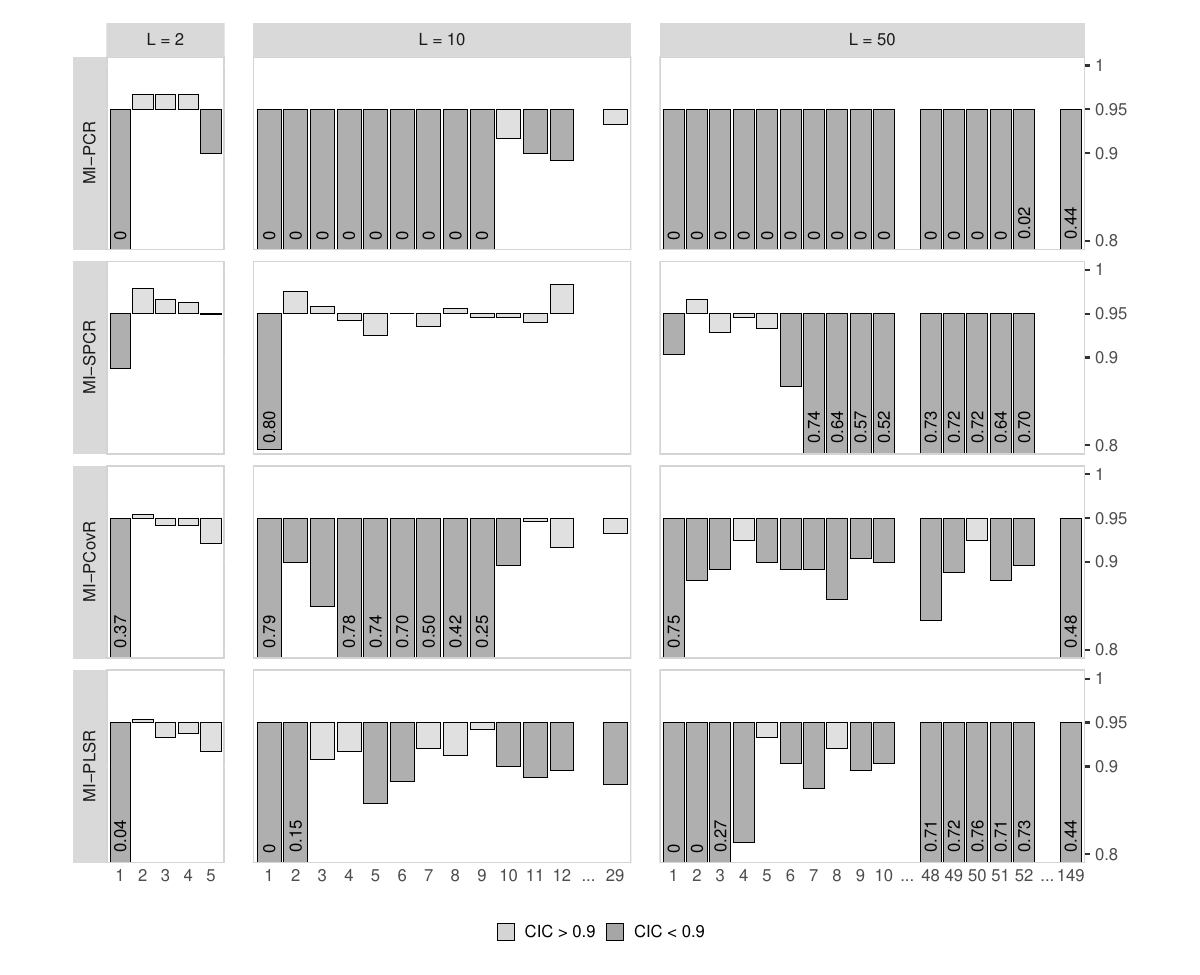} 

}

\end{knitrout}
  \caption{
    \label{fig:cic}
    The CIC for the estimated correlation coefficient between the first two items imputed is reported (Y-axis) as a function of the number of components ($nc$) used by the PCA-based imputation methods (X-axis).
    The plot is divided into a grid where the rows distinguish the results obtained after imputing the data with the four PCA-based methods and the columns distinguish the number of latent variables used to generate the data ($L$).
    All results plotted in this figure were obtained on data generated with \emph{mech} = MAR and $pm = 0.5$.
    For CIC values below 0.8, we reported the precise value within the corresponding bar.
  }
\end{figure}

In Figures~\ref{fig:prb},~\ref{fig:ciw}, and~\ref{fig:cic}, we report the PRB, CIW, and CIC for the correlation coefficient between $\mathbf{z}_1$ and $\mathbf{z}_2$ for different numbers of latent variables in the data-generating model ($L$) and numbers of components retained by the methods ($nc$).
Across all values of $L$, MI-PCR resulted in a smaller bias and coverage closer to nominal the more components were retained.
However, MI-PCR required the number of components to be greater or equal to the number of latent variables used in the data-generating model to return acceptable bias and coverages.
In particular, for $L = 2$ and $L = 10$, MI-PCR resulted in acceptable bias ($\text{PRB} < 10$) and close to nominal coverage ($\text{CIC} > 0.9$) only when using $nc \geq 2$ and $nc \geq 10$, respectively.
Contrary to expectation, this trend did not persist for all higher values of $L$ and $nc$.
For $L = 50$, MI-PCR resulted in high bias ($\text{PRB} > 20$) even for $nc = 50$.
Furthermore, for $L = 2$ and $L = 10$, MI-PCR resulted in large deviations from nominal coverage ($\text{CIC} < 0.9$) for $nc = 5$ and for $nc \in \{11, 12\}$, respectively.

Compared to MI-PCR, MI-SPCR performed much better, especially when using just a few components.
MI-SPCR resulted in the lowest bias, smallest confidence interval width, and closest to nominal coverage when using between 2 and 5 components.
For all values of $L$, using 2 components instead of 1, led to a large reduction in PRB and improvement in CIC.
Using 6 or more components had only a minor negative impact on the performance of the method in the condition with $L = 10$, resulting in a negligible increase in bias.
However, for $L = 50$, using 6 or more components did lead to high bias and low coverage.
Finally, the maximum number of components led to algorithmic failures, so there were no results to report for $nc = 29$ and $nc = 149$ in the $L = 10$ and $L = 50$ conditions, respectively.

MI-PCovR resulted in acceptable bias for all values of $L$, for all $nc$ values reported.
However, its bias performance was less stable than that of MI-SPCR across the values of $L$ and $nc$.
For $L = 10$, MI-PCovR led to smaller bias when using $nc = 2$ instead of $nc = 1$, but the bias increased when using $nc \in \{3, \dots, 9\}$, only to decrease again for $nc \geq 10$.
In the $L = 50$ condition, MI-PCovR resulted in decreasing bias for the range $nc \in \{1, \dots, 9\}$, and the lowest bias was achieved with $nc = L$, just after a small increase.
Furthermore, the CIC resulted in larger deviations from nominal coverage than MI-SPCR, resulting in acceptable CIC only with a few of the many $nc$ values considered.

Compared to MI-SPCR and MI-PCovR, the reduction in bias obtained by MI-PLSR for higher values of $nc$ was more gradual.
For $L = 10$ and $50$, using 2 components instead of 1, led to a reduction in bias, but 3 components were necessary to achieve $\text{PRB} < 10$, while both MI-SPCR, and MI-PCovR only needed 2 to achieve the same result.
However, the PRB remained small for $nc \in \{6, 7, 8, 9, 10\}$, even when that of MI-SPCR and MI-PCovR increased.
Despite this good bias performance, the CIW and CIC of MI-PLSR fluctuated between acceptable and not, with only a few values of $nc$ resulting in close-to-nominal coverage.

\begin{figure}
  \centering
\begin{knitrout}
\definecolor{shadecolor}{rgb}{0.969, 0.969, 0.969}\color{fgcolor}

{\centering \includegraphics[width=\maxwidth]{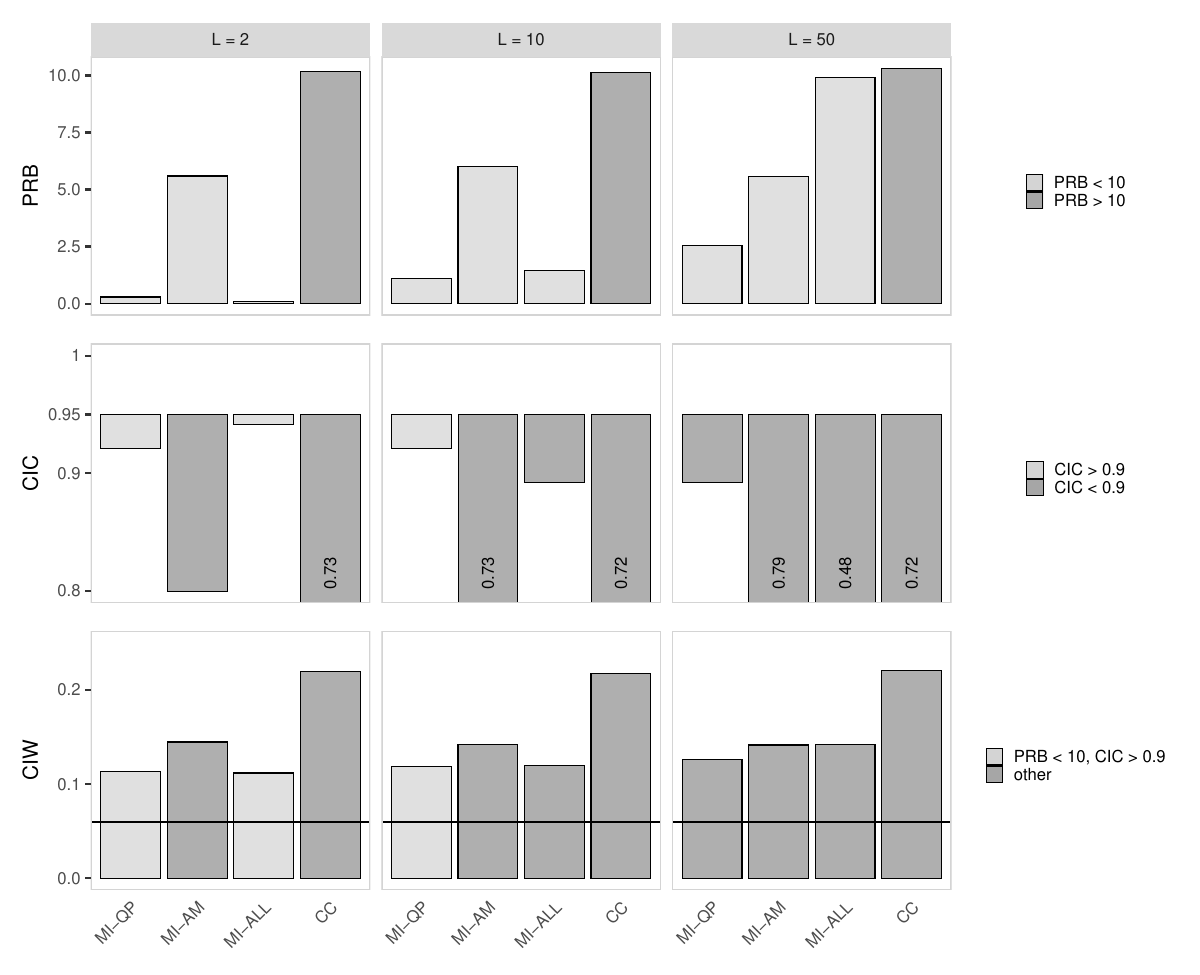} 

}

\end{knitrout}
  \caption{
    \label{fig:reference-methods}
    PRB, CIC, and CIW for the estimated correlation coefficient between the first two items imputed are reported for the traditional missing data handling methods considered.
    The black horizontal lines represent the average CIW obtained for the parameter of interest when analyzing the original fully observed data.
    For CIC values below 0.8, we reported the precise value within the corresponding bar.
  }
\end{figure}

To put these results into perspective, we reported the same performance metrics for three traditional MI approaches, and complete case analysis in Figure~\ref{fig:reference-methods}.
MI-QP and MI-ALL resulted in acceptable bias ($PRB < 10$) for all values of $L$, although larger values of $L$ did result in increased bias and decreasing coverage for both methods, but especially for MI-ALL.
MI-AM and CC did not result in a higher bias or lower coverages for larger values of $L$, but both returned relatively high bias and low coverage across all conditions.


\section{Discussion}\label{sec:discussion}

\subsection{Supervised dimensionality reduction}

Our simulation study outlined some clear advantages of using supervised dimensionality reduction techniques over standard PCA with MICE.
We found that MI-PCR requires the use of at least as many components as the number of latent variables in the data-generating model, which is in line with the results presented by~\cite{costantiniEtAl:2023}.
The simulation study presented here also showed that meeting this requirement is not sufficient to obtain good imputations for a large number of latent variables.
We found that the performance of MI-PCR does not only depend on knowing the number of latent variables in the data-generating model, but also on the number itself.
On the contrary, the SDR-based methods retaining just a few components resulted in small bias and good confidence interval coverage, \emph{independent} of the number of latent variables in the data-generating model.
Furthermore, when using any given number of components, except the maximum, using the SDR-based methods resulted in smaller bias, narrower confidence intervals, and closer to nominal coverage than MI-PCR.
Considering these results, SDR-based MICE seems to be more appropriate than PCR-based MICE for automatic imputation model specification.

Among the SDR-based methods, MI-SPCR had the best statistical properties.
MI-SPCR returned smaller bias and better coverage for a wider range of retained components compared to MI-PCovR.
MI-SPCR also achieved a smaller bias than MI-PLSR when retaining fewer components, and it resulted in consistently closer-to-nominal coverages.
Based on our results it seems that, at least in the context of the imputation of data with a latent structure, the \emph{discrete} type of supervision employed by MI-SPCR should be preferred to the \emph{continuous} supervision employed by MI-PCovR and MI-PLSR.

\subsection{Supervision and the number of principal components}

Based on our simulation study results, irrespective of which type of supervised dimensionality reduction is used, the implementation of SDR-based methods in MICE should aim for computing a small number of components.
Despite this general trend, MI-SPCR and MI-PCovR showed different performances in relation to the different numbers of components retained.

As described in the results section, the bias obtained by MI-SPCR was the smallest when using between 2 and 5 components, independently of the number of latent variables in the data-generating model, and it led to algorithmic failures for large numbers of components.
These results can be explained by considering how the number of PC retained influences MI-SPCR.
In MI-SPCR, supervision is introduced by pre-screening the columns of possible predictors set to exclude any predictors that are not correlated strongly enough with the variable under imputation. 
By reducing the number of columns in the predictor set, this supervision reduces the maximum number of components that can be estimated.
The inverse constraint also holds. 
Fixing the number of components puts an upper-bound on the number of variables that can be excluded during the screening process.
So, for example, when using 50 components, MI-SPCR must retain, at least, 50 predictors during the screening step, regardless of how weakly some of these variables may associate with any given variable under imputation.
Retaining more components forces the threshold values to be smaller and results in keeping more predictors that are less strongly related to the dependent variable, and, by doing so, it limits the advantage of using supervision in the PCA.
In conclusion, considering our results and the relationship between the number of components and supervision, we recommend retaining between 2 and 5 components when using MI-SPCR.

In the results section, we noted that in the condition with 10 latent variables, MI-PCovR led to smaller bias when using 2 components instead of 1, but its bias increased when retaining 3 to 9 components, only to decrease again when retaining 10 or more components.
A similar but less extreme trend was also detected in the condition with 50 latent variables, a result which we explore further in the appendix.
This fluctuating bias performance can be explained by considering the relationship between the number of components retained and the way we computed $\alpha$ in our simulation study.
As described in equation~\ref{eq:aml}, for the same data set, $\alpha_{ML}$ is bigger when the unexplained variance by the components retained is smaller.
The more components we retained in our simulation study, the closer the value of $\alpha$ was to 1, and the closer MI-PCovR became to MI-PCR.
As a result, MI-PCovR resulted in smaller bias than MI-PCR when retaining the first components, as supervision helped to compute leading components that were important predictors for the imputation task.
However, as more components were retained, the effect of supervision started to diminish, which drove the bias closer to that of MI-PCR.
After this initial increase, the bias achieved by MI-PCovR dropped when retaining as many components as the number of latent variables, mirroring the drop in bias presented by MI-PCR for the same number of components.
As a result, the best performances for MI-PCovR could be achieved by retaining the first components or by retaining a number of components just above the number of latent variables.
Because of this fluctuating performance, the optimal range of components to consider when using MI-PCovR is not as clear as for MI-SPCR.

\section{Limitations and future directions}\label{sec:limits}

An important aspect to consider in deciding which version of supervised dimensionality reduction to use with MICE is how flexible these approaches are to deviations from normality of the data.
This topic was not covered by our simulation study.
MI-SPCR can easily be adapted to impute binary and categorical variables, and the only complication would be in defining a suitable threshold parameters.
One option would be to estimate the associations via fit measures derived from simple (multinomial) logistic regression models.
Much research has been dedicated to extending PLSR to categorical outcomes \citep[e.g., ][]{dingGentleman:2005, chungKeles:2010}, and these approaches could be used in a similar way to the standard PLS implementation we used in this study.
The development of PCovR for classification tasks has not received much attention \citep[][is the only example of which we are aware]{parkEtAl:0000}, but the same approaches used to fit PLS in the generalized linear framework should also apply to PCovR.
However, the maximum likelihood estimation of the $\alpha$ parameter can only be done for continuous dependent variables.
To impute categorical variables, PCovR would require cross-validation to estimate the value of $\alpha$, adding to the computational intensity of the procedure.

The set of predictors used to compute the components can also include categorical variables.
There are different ways of accommodating these categorical variables when estimating the components, including the naive application of traditional PCA~\cite{filmerPritchett:2001} and the PCAMIX algorithm \citep[][]{kiers:1991, chaventEtAl:2012, chaventEtAl:2017} specifically designed for this purpose.
Which of these approaches is more appropriate for the predictive task involved in MICE is yet to be tested.

Finally, all of the PCA-based methods considered here (both supervised and unsupervised) entail a high computational load.
For every variable and every iteration of the MICE algorithm, complex matrix operations need to be performed to estimate the components.
When cross-validating the tuning parameters, the supervised approaches described in this article can increase.
Future research should explore possible computational shortcuts to perform supervised dimensionality reduction faster~\citep[e.g., ][]{abrahamInouye:2014, halkoTropp:2011}.

\section{Conclusions}\label{sec:conclusions}

Based on the simulation study presented here, it can be concluded that adding a supervision element to the classical use of PCR as a univariate imputation method can improve significantly the performance of MI-PCR, especially when the data contain hundreds of variables.
Although there is room to assess the performance of these imputation methods in more complex data scenarios, MI-SPCR was particularly effective for the imputation of missing values and seems to be preferable to MI-PCovR and MI-PLSR.

\setcounter{secnumdepth}{4}

\section{Code availability}

The R code used to perform the simulation study is available on Zenodo \citep{costantini:2022c}.
Please read the README.md files for instructions on how to replicate the results.
The article is also accompanied by an interactive results dashboard packaged as an R Shiny app \citep{costantini:2022b}.
We encourage the interested reader to use this tool while reading the results and discussion sections.
A user manual is included as a README file in the folder accessible through the DOI provided in the citation.
The software can be downloaded and installed as an R package.

\bibliography{\pathBIB/bibshelf}


\section{Appendix: Further thoughts on MI-PCovR and MI-PCR}\label{sec:pcovr-pcr}
\appendix



We noted in Section~\ref{subsec:meth-pcovr} how PCR can be seen as the special case of PCovR when $\alpha = 1$.
In our simulation study, the value $\alpha$ was defined by Equation~\ref{eq:aml}, given a certain number of PCs.
The more PCs were retained in MI-PCovR, the lower the value of $\hat{\sigma}_{\mathbf{E}_{\mathbf{X}}}$ was, and the higher the value of $\alpha$ became.
This characteristic was reflected by the tendency of MI-PCovR to converge to the performance of MI-PCR the more PCs were retained.

\begin{figure}
    \centering
\begin{knitrout}
\definecolor{shadecolor}{rgb}{0.969, 0.969, 0.969}\color{fgcolor}

{\centering \includegraphics[width=\maxwidth]{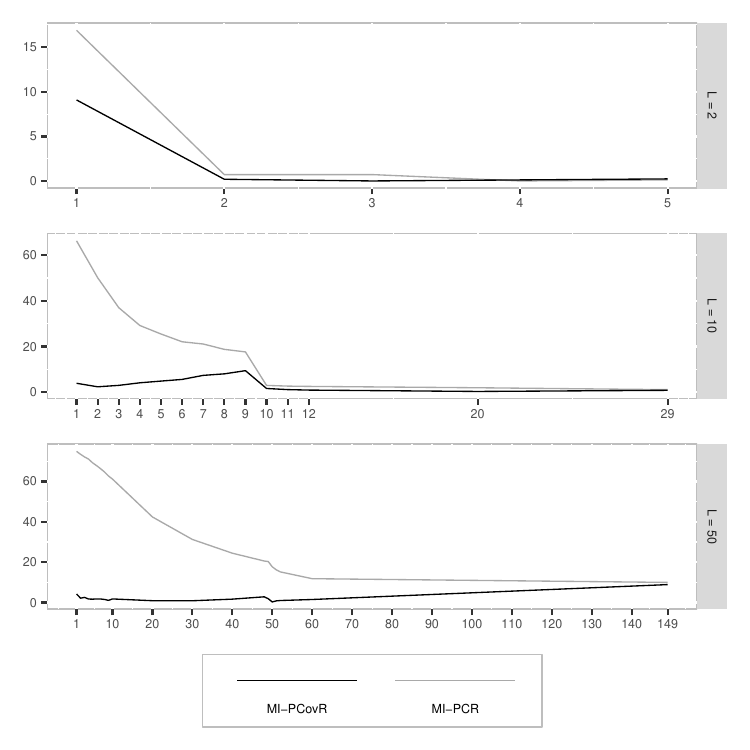} 

}

\end{knitrout}
    \caption{
      \label{fig:pcovr-pcr}
      PRB (Y-axis) for the correlation between $\mathbf{z}_1$ and $\mathbf{z}_2$ as a function of $nc$ (X-axis) obtained by imputing the data with MI-PCR and MI-PCovR.
      The data generation condition reported is $mech$ = MAR, $pm = 0.5$.
    }
  \end{figure}

In Figure~\ref{fig:pcovr-pcr}, we report the bias trends for estimating the correlation between $\mathbf{z}_1$ and $\mathbf{z}_2$, after imputing the data with MI-PCR and MI-PCovR, as a function of a wider range of components compared to what reported in Section~\ref{subsec:results}.
For $L \in \{2, 10, 50\}$, MI-PCR gradually resulted in smaller bias for higher values of $nc$, but a steep change in the performance of MI-PCR resulted from setting $nc = L$.
MI-PCovR had a smaller bias than MI-PCR for $nc = 1$ for all $L \in \{2, 10, 50\}$.
However, the PRB obtained with MI-PCovR increased as $nc$ increased for $L \in \{10, 50\}$, but, similarly to MI-PCR, it dropped for $nc = L$.
For $nc > L$, the PRB gradually increased again and its performance converged to that of MI-PCR as $nc$ approached its maximum.
This is in line with the understanding of PCR as a special case of PCovR, and it confirms that the more components are retained by MI-PCovR, the closer we can expect its performance to be to that of MI-PCR.

\end{document}